# Magnetic measurements and $^{57}$Fe Mössbauer spectroscopy in oxygen deficient SmFeAsO$_{0.85}$

Israel Felner, Israel Nowik and Menachem I. Tsindlekht

Racah Institute of Physics, The Hebrew University, Jerusalem, 91904, Israel

Zhi-An Ren, Xiao-Li Shen, Guang-Can Che and Zhong-Xian Zhao

National Laboratory for Superconductivity, Institute of Physics and Beijing National Laboratory for Condensed Matter Physics, Chinese Academy of Sciences, P. O. Box 603, Beijing 100190, P. R. China

Magnetic measurements and $^{57}$Fe Mössbauer spectroscopy studies were performed on oxygen-deficient high temperature superconductor SmFeAsO$_{0.85}$ with $T_C$=52.4 K. The upper-critical behavior ($H_{C2}$) values were extracted from the real part of ac measurements. The field dependence of $H_{C2}$ is consistent with a two band model. Mössbauer spectra below and above $T_C$ consist of a singlet and a doublet, which are attributed to Fe ions which have two or one oxygen ions in their close vicinity, respectively. No change is observed in the major (~75%) singlet related to Fe ions surrounded by two oxygen ions. On the other hand, the doublet which senses oxygen vacancies shows a well defined magnetic sextet below $T_C$. This indicates *coexistence* on a microscopic level of the two mutually exclusive states namely: superconductivity which is confined to the Fe-As layers and magnetism, in the same layers. Alternatively, the hyperfine parameters of the doublet are similar to the reported values of FeAs which orders magnetically at 77 K. Thus the magnetic features observed below $T_C$, may be related to FeAs as an extra phase.

**Introduction**
The recent discovery of high temperature superconductivity (HTSC) in the layered Fe-based RFeAs(O,F) (R=rare-earth) with $T_C$>50 K has triggered an intense research in this oxy-pnictides system.[1] The RFeAs(O,F) system has a layered crystal structure with alternating Fe-As and R-O sheets, where the Fe atoms are arranged on a simple square lattice and each Fe has two oxygen ions in its close neighbor environment. Superconductivity (SC) is confined to the Fe-As layers, in contrast to the well-known HTSC systems in which SC are confined to the Cu-O sheets. Beside the high $T_C$, there are rather further striking similarities to the Cu-O based HTSC compounds. In both systems competition between superconductivity (SC) and magnetism exist. SC emerges when doping a magnetic parent compounds with holes (or electrons) is done and thereby suppressing the magnetic order. However in these pnictides HTSC, it is achieved upon electron doping and not by hole doping as in most of the cuprates.

As stated above, the un-doped parent compounds RFeAsO (R=La, Ce, Nd, Sm and Gd) are not SC and neutron measurements for LaFeAsO, have revealed a commensurate spin-density-wave (SDW) order of the Fe moments below ~150 K with an amplitude of 0.35$\mu_B$.[2-3] This result was confirmed by $\mu$SR[4] and Mossbauer studies.[5] This suggests that spin fluctuations play an important role in the pairing mechanism, similar to the cuprates HTSC compounds. Previous experiments indicated that F-doping at the oxygen site, suppress the magnetic nature of the system and further doping induce SC with $T_C$>50 K.[6-7] From this point of view, producing oxygen vacancies would be equivalent to F$^-$ doping. Indeed, the Chinese group of the present paper has succeeded in preparing RFeAsO$_{1-\delta}$ superconductors by high pressure synthesis. By tuning the oxygen content they were able to reduce the magnetic transition of the parent compound, to produce SC and to construct a full phase diagram.[8]

Here we report dc and ac magnetization studies on the SC SmFeAsO$_{0.85}$ (T$_C$~52.4 K) from which the magnetic critical field (H$_{C2}$) values are deduced. $^{57}$Fe Mossbauer spectroscopy (MS) studies below and above T$_C$, reveal two Fe sites, their intensity ratio fits precisely the oxygen vacancies concentration. The intense singlet (75%) corresponds to those iron nuclei which do not have a vacancy in their immediate neighborhood. The second sub-spectrum corresponds to Fe nuclei which have vacancies in their close vicinity and shows a quadrupole doublet at 95 K. Surprisingly at 4.2 K, this doublet disappears and instead, a well-defined magnetic sextet is observed. This implies that SC (which is confined to the Fe-As layers) and magnetism coexist in the same Fe-based layer. It is the first experiment, which probes **directly** the coexistence of these two states. The phase separation option is excluded.

**Experimental Details.**

The SmFeAsO$_{0.85}$ SC (assigned as S4192) with the optimal oxygen concentration for higher T$_C$ material was prepared by high pressure synthesis.[9] SmAs (pre-sintered), Fe and Fe$_2$O$_3$ were mixed together according to the nominal composition, then ground and pressed into pellets which were sealed in a BN crucible and sintered under 6 GPa at 1250 C for two hours. The purity and the tetragonal (S.G. P4/nmm) structure (a=3.897(6) A$^o$ and c=8.407(1) A$^o$) were verified by powder x-ray diffraction (XRD) MXP18A-HF type diffractometer. Zero-Field-Cooled (ZFC) and field-cooled-cooling (FCC) dc magnetic measurements in the range of 5-200 K were performed in a commercial (Quantum Design) super-conducting quantum interference device (SQUID) magnetometer. The ac susceptibility was measured by a home-made probe inserted in the SQUID magnetometer, with excitation frequency and amplitude of 115 Hz and 50 mOe respectively. The Mössbauer studies were performed using a conventional constant acceleration drive and a 50 mCi $^{57}$Co:Rh source. The velocity calibration was performed using a room temperature α-Fe absorber, the isomer shifts (I.S.) values reported are relative to that of iron. The observed spectra were least square fitted by theoretical spectra, assuming a distribution of hyperfine interaction parameters, corresponding to in-equivalent iron locations differing in local environment.

**Experimental Results.**

*(a) dc and ac magnetic measurements*

Comprehensive dc and ac magnetic measurements have been performed on SmFeAsO$_{0.85}$, and a selection of the most prominent magnetic features obtained, are presented in Figures 1-4. Prior to each cooling to 5 K and recording the M$_{ZFC}$ signal or the *ac* measurement, the SQUID magnetometer was adjusted to be in a *"real"* H=0 state. The ZFC and FCC magnetization curves measured at 5 Oe are presented in Figure 1. The FCC branch was taken via cooling the samples under 5 Oe down to 5 K. The deviation from the straight line at T$_C$= 52.4 K is our definition for the onset of the SC state. Above T$_C$, the M(T) curve measured at 2.5 kOe (Figure 2, inset) does not follow the Curie-Weiss law and shows a linear behavior, a typical trend for Sm$^{3+}$ (Van Vleck ion) found in many Sm-based compounds.[10] Isothermal magnetization measurements have been performed at various temperatures. Generally speaking, below T$_C$, the M(H) plots exhibit the typical SC shape, however, the paramagnetic Sm contribution does not permit an easy determination of the upper critical field (H$_{C2}$). For that purpose the ac susceptibility studies have been used as presented in Figures 3-4.

Figure 2 shows an almost linear M(H) behavior at 70 K. The extrapolation of the linear part to H=0 yields a small remnant moment (0.0026 emu/g) which is equivalent to an upper limit of 11 ppm of Fe metal as an impurity phase. The paramagnetic effective moment deduced from the slope, yields P$_{eff}$= 0.82μ$_B$, a value which is very close to the expected value 0.85μ$_B$ for Sm$^{3+}$ ions. That indicates that the Fe ions do not contribute to the magnetic moment at elevated temperatures.

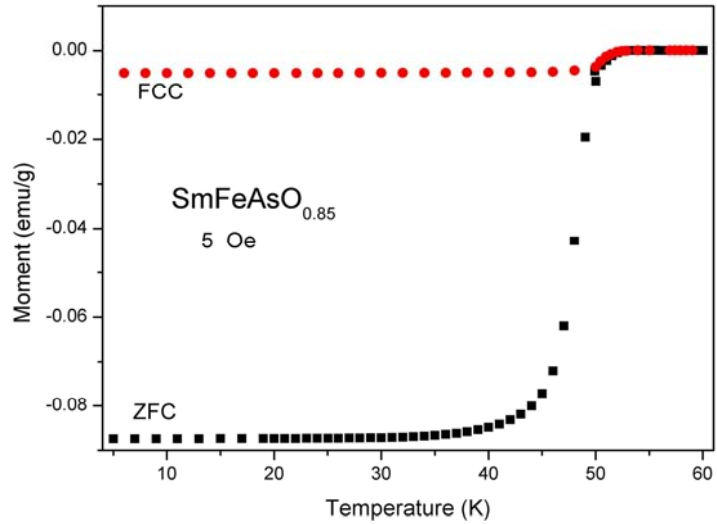

Figure 1. ZFC and FCC magnetization curves of SmFeAsO$_{0.85}$ measured at 5 Oe.

*AC* susceptibility measurements have been performed under various applied fields and some representative real and imaginary parts of the susceptibility are presented in Figure 3. The shift of the signals with H toward low temperatures is readily observed. Strong vortex pinning, and the particulars of the grain disorientation distribution, complicates the interpretation of H$_{C2}$ in polycrystalline materials. We define H$_{C2}$ as the onset in the real parts in Figure 3, and the temperature dependence of H$_{C2}$ is depicted in Figure 4. As is evident, H$_{C2}$(T) exhibits a non continuous trend. For low H values a significant upward curvature is observed, whereas for H> 1.5 T, H$_{C2}$ is linear with temperature. The different behavior of H$_{C2}$(T) at low and at high applied fields, may suggest a two-band SC in the pnictides, as predicted by several theoreticians[11-13] and found experimentally in F$^-$ doped LaFeAsO$_{0.89}$F$_{0.11}$ sample.[14-15] By extrapolation of the linear part to T=0, we obtained H$_{C2}$(0) ~ 45 T, a value which is with fair agreement with Ref. 14, but much lower than 150 T extracted from specific-heat measurements.[15]

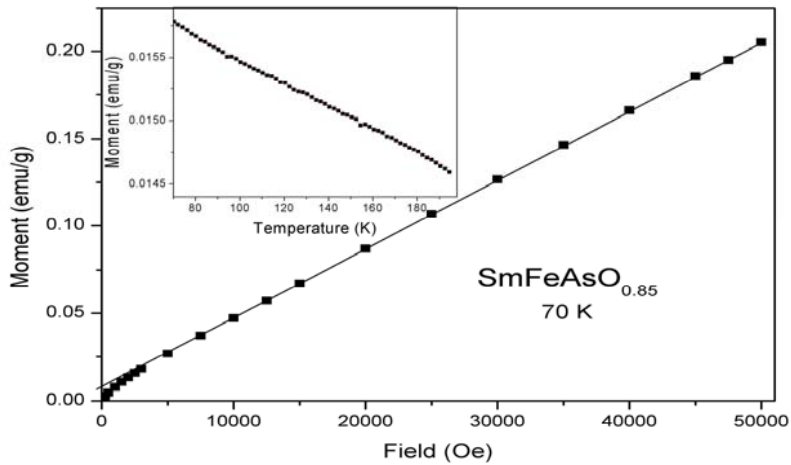

Figure 2. Isothermal magnetization of SmFeAsO$_{0.85}$ measured at 70 K. The temperature dependence of the magnetization measured at 2.5 kOe is shown in the inset.

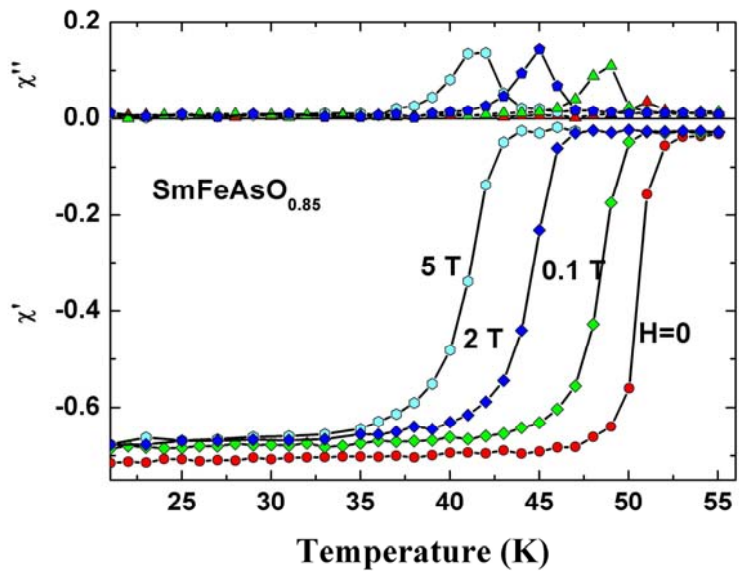

Figure 3. Temperature dependence of real and imaginary parts of the ac susceptibility (arb. units) of SmAsFeO$_{0.85}$ measured under various applied dc fields.

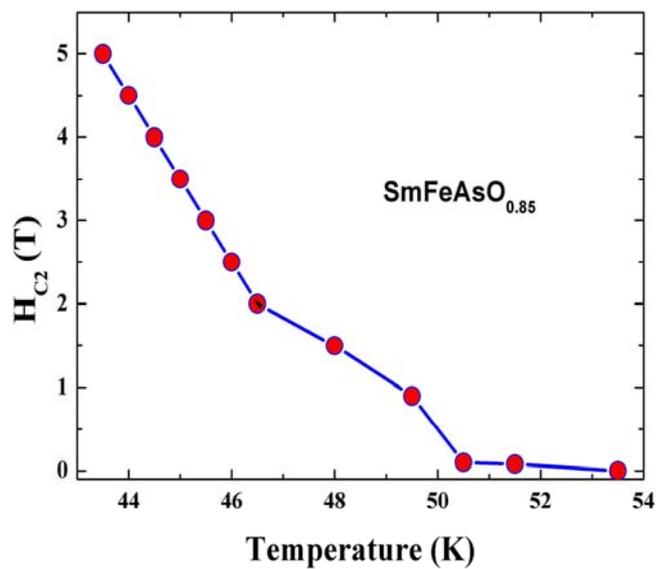

Figure 4. The temperature dependence of H$_{C2}$, indicating two bands structure

(b) *$^{57}$Fe Mössbauer measurements*
Our main interest here is the Mössbauer spectroscopy (MS) studies performed on SmAsFeO$_{1-x}$ (x=0.15) measured at 4.2 and 95 K shown in Figure 5 and the hyperfine interaction parameters obtained are given in Table 1. In the tetragonal P4/nmm structure each Fe which resides in the (2b) crystallographic position has two oxygen ions as nearest neighbors along the crystal c-axis with the shortest Fe-O distance of c/2=4.2035 A. Each spectrum in Figure 5 is composed of two sub-spectra. The intense sub-spectrum (~75%) is almost a singlet, and corresponds to those iron nuclei which have two oxygen ions in their close vicinity. The second sub-spectrum, a quadrupole doublet at 95 K and a magnetic sextet at 4.2 K, corresponds to iron nuclei, which have vacancies in their immediate neighborhood. In the case of random distribution of vacancies, the probability of an iron ion (in this layered structure) to have one or two vacancies as first nearest neighbor is $2x(1-x)$ and $x^2$, respectively. For x=0.15 the fraction of iron nuclei with a distorted environment, and thus displaying large quadrupole splitting, is 25.5% for those with single neighboring vacancy and 27.75% when iron with two neighboring vacancies are also included. Experimentally, Table 1 shows the relative intensity of the two subspectra. Indeed, 25(2)% of the spectral area displays local distortions at 95 K and 27(1)% at 4.2 K, indicating that the vacancies fraction is in fact extremely close to x=0.15. The isomer shift values for both sites are typical to a divalent low-spin state for all Fe ions. This is consistent with the experimental observation (Figure 2) that Fe does not contribute to the magnetic moment at elevated temperatures.

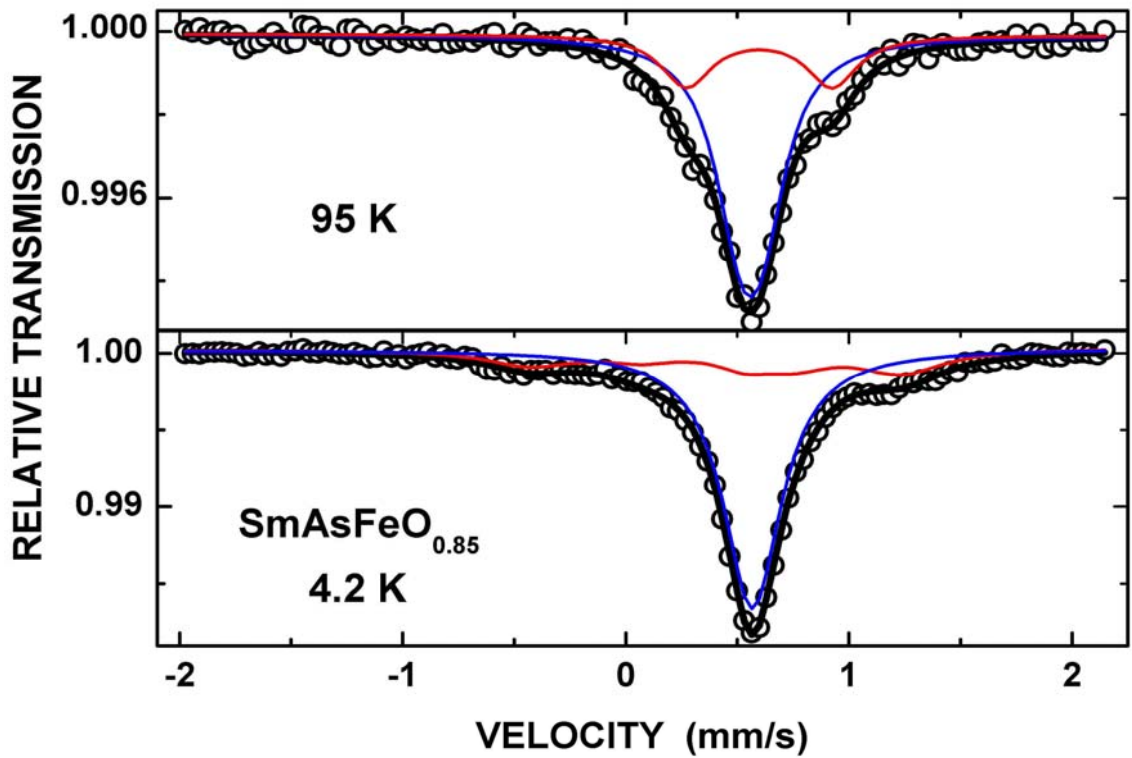

Figure 5. The Mössbauer spectra of $SmAsFeO_{0.85}$ measured at 4.2 and 95 K.

Figure 5 and Table 1 show that the hyperfine parameters of the major sub-spectrum are identical at both temperatures and are very similar to values obtained in other SC compounds such as $LaAsFeO_{0.89}F_{0.11}$[5] and LaFePO.[16] However the most important observation is that there are large changes in the hyperfine interaction parameters of the Fe nuclei with distorted environments. Above $T_c$ a well-defined quadrupole doublet is obtained, whereas at 4.2 K a magnetic sextet is developed. The later sub-spectrum was analyzed in terms of diagonalization of the full spin Hamiltonian in which $\Theta$ is the angle between the electric field gradient axial c-axis and the magnetic hyperfine field ($H_{eff}$). $H_{eff}$ acting on the Fe nuclei is 40(1) kOe, similar to the low fields observed for the non-SC magnetic LaAsFeO compound.[5] The angle $\Theta$ turns out to be 86(8)° indicating that $H_{eff}$ and hence the iron magnetic moments lie in the basal plane. This result is consistent with neutron diffraction studies on LaAsFeO in which an antiferromagnetic structure below $T_N$=134 K is deduced and Fe moment direction is within the Fe-planes.[20] It seems that the vacancy environment is strongly changing at the SC transition, sensed by the close iron nuclei, which become **magnetically** ordered. To our best knowledge, this is the first case in which such drastic changes are observable by Mössbauer spectroscopy at the superconducting transition. Our major statement is as follows: in Figure 5 we definitely observe a real coexistence of both superconductivity and magnetic order in the same iron planes.

The assumption of a phase separation of two SC and magnetic phases is completely ruled out. In the pnictides, SC is induced by oxygen vacancies, since the stoichiometric RAsFeO (x=0) compounds show a SDW behavior and are not SC. A separate magnetic phase due to oxygen vacancies would not induce SC in the rest of the material.

The MS spectra of the two SC LaAsFeO$_{0.89}$F$_{0.11}$[5] and LaFePO[16] samples, exhibit only one singlet with hyperfine parameters similar to the singlet parameters listed in Table 1. In these samples no vacancies exist and therefore no second sub-spectrum is achieved. The vacancies obtained in our sample induce two types of Fe ions, and thus permits a direct insight into the various Fe ions behavior. μSR studies indicate a complex of magnetic correlations in the SC state of SmAsFeO$_{0.82}$F$_{0.18}$,[4] but no such correlations were observed in LaFeAsO$_{1-x}$F$_x$.[17] Therefore, the magnetic correlations[4] were attributed to the coupling between the Sm and Fe moments, which is absent in the La based compound. The advantage of the present study is that MS probes precisely all the various $^{57}$Fe ions, which specify the magnetic state of those Fe ions which experience vacancies in their close vicinity.

| T (K) | INT(1) (%) | GAM mm/s | I.S. (1) mm/s | EQ(1) mm/s | INT(2) (%) | I.S.(2) mm/s | EQ(2) mm/s | $H_{eff}$ (kOe) | $\Theta$ ° |
|---|---|---|---|---|---|---|---|---|---|
| 95 | 74 | 0.27(2) | 0.560(3) | 0.05(1) | 26(2) | 0.60(1) | 0.33(1) | 0 | --- |
| 4.2 | 73 | 0.31(1) | 0.568(1) | 0.01(1) | 27(1) | 0.53(1) | 0.41(2) | 40(1) | 86(8) |

Table 1. The hyperfine interaction parameters for SmAsFeO$_{0.85}$ derived from the analysis of the Mössbauer spectra. INT, GAM, I.S., EQ and $H_{eff}$ stand for relative intensity, line width, isomer shift, quadrupole parameter (EQ=eqQ/4) and magnetic hyperfine field. $\Theta$ is the angle between the electric field gradient axial axis and $H_{eff}$.

Coexistence of the two mutually exclusive SC and magnetic states have been observed in the two layered Cu-O based HTSC systems: RuR$_{2-x}$Ce$_x$Sr$_2$Cu$_2$O$_{10-\delta}$[18] and RuSr$_2$RCu$_2$O$_8$.[19] In the ruthenates, the SC state originates from the CuO$_2$ planes, whereas the magnetic state is confined to separate Ru layers, which are practically decoupled from the CuO$_2$ planes, so that there is no pair breaking. On the other hand, in SmAsFeO$_{0.85}$ the two states originate from the same Fe-As layers, thus provide the first example for a real coexistence between them. The current state of experiments does not allow us to suggest any consistent model for the coexistence phenomenon presented here, which definitely warrants further theoretical and experimental investigations.

Alternatively, our preferred scenario is that the hyperfine parameters for the minor sub-spectra listed in Table 1, namely the doublet observed at 95 K and the magnetic splitting at 4.2 K, are similar to the reported hyperfine parameters of FeAs, [21] which order magnetically at 77 K[22]. Therefore the magnetic splitting observed in Fig. 5, may be attributed to the existence FeAs as an extra phase. The complex magnetic correlations in the SC state observed in SmAsFeO$_{0.82}$F$_{0.18}$ by μSR studies[4] may also be explained by the presence of this extra phase. MS measurements as well as μSR studies on single crystal samples are urgently needed in order to clarify the origin for the magnetic correlations below T$_C$ .

**Acknowledgments**: This research is partially supported by the Israel Science Foundation (ISF, 2004 grant number: 618/04) and by the Klachky Foundation for Superconductivity.